\documentclass[reprint, superscriptaddress, amsmath, amssymb, aps, physrev]{revtex4-2}

\usepackage{graphicx}
\usepackage{xcolor}
\usepackage[percent]{overpic}
\usepackage{subfigure}
\usepackage{tikz}
\usepackage{braket}
\usepackage{hyperref}

\usepackage{helvet}

\usepackage{sfmath}
\usepackage[font=sf, labelfont=sf, justification=raggedright, singlelinecheck=false]{caption}

\usepackage[normalem]{ulem}
\newcommand{\new}[1]{\textcolor{blue}{#1}}
\newcommand{\old}[1]{\textcolor{purple}{\sout{#1}}}

\renewcommand{\new}[1]{#1}
\renewcommand{\old}[1]{}

\begin{document}

\title{First-Principles Evidence for Strongly Correlated Superconductivity Driven by Structural Variations in \texorpdfstring{La$_3$Ni$_2$O$_7$}{La3Ni2O7}}

\author{Daan Verraes}
\affiliation{Center for Molecular Modeling, Ghent University, Technologiepark 46, 9052 Zwijnaarde, Belgium}
\affiliation{Department of Physics and Astronomy, Ghent University, Krijgslaan 281, 9000 Ghent, Belgium}

\author{Tom Braeckevelt}
\affiliation{Center for Molecular Modeling, Ghent University, Technologiepark 46, 9052 Zwijnaarde, Belgium}

\author{Nick Bultinck}
\email{e-mail: Nick.Bultinck@UGent.be}
\affiliation{Department of Physics and Astronomy, Ghent University, Krijgslaan 281, 9000 Ghent, Belgium}

\author{Veronique Van Speybroeck}
\email{e-mail: Veronique.VanSpeybroeck@UGent.be}
\affiliation{Center for Molecular Modeling, Ghent University, Technologiepark 46, 9052 Zwijnaarde, Belgium}

\begin{abstract}
We conduct first-principles simulations of \texorpdfstring{La$_3$Ni$_2$O$_7$}{La3Ni2O7}, a nickelate in which recent experiments have shown signs of high-temperature superconductivity. Within the hydrostatic pressure range where superconductivity is observed, we find a significant increase in effective on-site repulsion in the maximally localised Wannier functions comprising the Ni $e_g$ bands crossing the Fermi energy. We attribute this increase to an interplay between orbital localisation and competing screening channels arising from structural variations. Our results indicate that the superconducting region in the \texorpdfstring{La$_3$Ni$_2$O$_7$}{La3Ni2O7} phase diagram coincides with a region of enhanced electronic correlations, which show a close correspondence with the critical temperature. Including finite temperatures up to 100 K, \textit{ab initio} molecular dynamics simulations then provide new insights into the debated structural phase diagram and further clarify the origin of the right-triangular superconducting dome. Finally, we study \texorpdfstring{Ac$_3$Ni$_2$O$_7$}{Ac3Ni2O7} to confirm the crucial role of the \textit{A}-site cation in shaping the pressure-driven evolution of electronic correlations.
\end{abstract} 

\makeatletter
\def\@fnsymbol#1{*}
\makeatother
\maketitle

\setcitestyle{super}

Nearly four decades after the discovery of high-temperature (high-$T_\text{c}$) superconductivity in hole-doped cuprates, a recent breakthrough occurred with the observation of superconductivity up to $80$ K in bulk samples of the nickelate La$_3$Ni$_2$O$_7$ under hydrostatic pressures above 14 GPa \cite{bednorz_possible_1986, sun_signatures_2023, zhang_high-temperature_2024}. Subsequent transport measurements suggest a right-triangular superconducting region in the pressure-temperature (PT) phase diagram between 14 and 80 GPa, with a maximum $T_\text{c}$ at 18 GPa \cite{li_identification_2025}. The structural and electronic differences with the cuprates have thereby motivated a plethora of theoretical works to unmask the pairing mechanism at play \cite{zhang_electronic_2023, luo_bilayer_2023, gu_effective_2023, luo_high-tc_2024, yang_possible_2023, lechermann_electronic_2023, shen_effective_2023, sakakibara_possible_2024, lu_interlayer_2024, liao_electron_2023, qu_bilayer_2024, yang_interlayer_2023, wu_superexchange_2024, huang_impurity_2023, jiang_high-temperature_2024, lu_superconductivity_2023, oh_type_2023, zhang_structural_2024,kaneko_pair_2024, YangHui_2024, YangHuiOh_2024, shilenko_correlated_2023, tian_correlation_2024, liu_-wave_2023, cao_flat_2024, qin_high-T_c_2023, chen_charge_2024, jiang_pressure_2024, christiansson_correlated_2023, zhang_trends_2023, yi_antiferromagnetic_2024, chen_orbital-selective_2024, jiang_theory_2025}.

\begin{figure}[t]
    \centering
    \includegraphics[width=\linewidth]{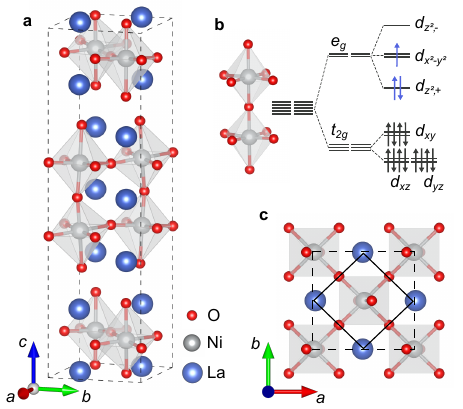}
    \caption{\textbf{Crystal structure and low-energy configuration of La$_3$Ni$_2$O$_7$.} 
    \textbf{a} Orthorhombic unit cell of the $Cmcm$ phase, characterized by tilted NiO$_6$ octahedra, shown in gray.
    \textbf{b} Crystal field splitting and orbital-selective $e_g$ configuration (blue) of two Ni$^{2.5+}$ cations in a dimer. The $d_{z^2}$ orbitals form bonding $d_{z^2,+}$ and anti-bonding $d_{z^2,-}$ states. 
    \textbf{c} Top view of a NiO$_2$ bilayer. Solid and dashed black frames represent the $Fmmm$ and $Cmcm$ unit cells, respectively.}
    \label{fig: introduction}
\end{figure}

As a member of the Ruddlesden-Popper nickelates, the conventional polymorph of La$_3$Ni$_2$O$_7$ features intact bilayers of NiO$_6$ octahedra (Fig.~\ref{fig: introduction}a), with a formal nickel oxidation of Ni$^{2.5+}$ ($3d^{7.5}$). This hole concentration indicates a multi-orbital low-energy configuration, which differs significantly from the Ni$^{+}$ ($3d^{9}$) cations forming the square-planar NiO$_2$ layers in infinite-layer nickelates, or from the Cu$^{2+}$ ($3d^{9}$) cations of the CuO$_2$ planes in cuprates \cite{li_superconductivity_2019, bednorz_possible_1986}. Density functional theory (DFT) calculations indicate that the electron density near the Fermi energy ($E_\text{F}$) is comprised mostly of the Jahn-Teller distorted Ni $e_g$ orbitals, which exhibit hybridisation with adjacent O $2p$ orbitals along their symmetry-aligned axes \cite{luo_bilayer_2023, gu_effective_2023, zhang_electronic_2023}. These studies suggest that the active Ni $d_{z^2}$ orbitals form strong $\sigma$-bonds with the apical O $p_z$ orbitals connecting the bilayers along the $c$-axis in Fig.~\ref{fig: introduction}a, giving rise to a large interlayer hopping which splits the $d_{z^2}$ orbitals into bonding and anti-bonding states, while the $d_{x^2-y^2}$ orbitals remain approximately decoupled between monolayers. This orbital-selective (2Ni)$^{5+}$ dimer configuration, shown in Fig.~\ref{fig: introduction}b, suggests the construction of an effective model comprising the four $e_g$ orbitals for which many bilayer Hubbard and \textit{t-J} models have been proposed and studied in literature \cite{luo_bilayer_2023, luo_high-tc_2024, gu_effective_2023, yang_possible_2023, lechermann_electronic_2023, shen_effective_2023, sakakibara_possible_2024, lu_interlayer_2024, liao_electron_2023, qu_bilayer_2024, yang_interlayer_2023, wu_superexchange_2024, huang_impurity_2023, jiang_high-temperature_2024, lu_superconductivity_2023, oh_type_2023, zhang_structural_2024, kaneko_pair_2024,YangHui_2024, YangHuiOh_2024}.

On the experimental side, the crystal structure of La$_3$Ni$_2$O$_7$ at superconducting conditions is not yet completely resolved due to resolution constraints in the reported X-ray diffraction (XRD) data. At room temperature, the bilayer polymorph exhibits a first-order structural transition between two orthorhombic ($a \neq b$) phases, symmetrizing from space group $Cmcm$ to subgroup $Fmmm$ above a hydrostatic pressure of 10 GPa \cite{sun_signatures_2023}. A key structural change at the transition is the $c$-axis alignment of NiO$_6$ octahedra, depicted in Fig.~\ref{fig: introduction}c. This octahedral alignment coincides with a metallisation of the $\sigma$-bonding $d_{z^2}$ states, which has been suggested to stabilise bilayer superconductivity at low temperatures \cite{sakakibara_orbital_2014}. Subsequent \textit{in situ} synchrotron XRD identified the high-pressure phase at a much lower temperature of $\sim 40$ K to be a third, tetragonal ($a=b$) phase with space group $I4/mmm$ \cite{wang_structure_2024}, in agreement with DFT predictions at 0 K \cite{geisler_structural_2024}. However, the precise boundaries between these structural phases in the complete pressure-temperature phase diagram remain to be located.

In this work, we conduct first-principles calculations of La$_3$Ni$_2$O$_7$ to uncover its structural phase diagram and reveal how the experimentally observed right-triangular superconducting region emerges from it. We first perform full-cell enthalpic optimisations under varying hydrostatic pressure to investigate the crystal structure at zero temperature and examine how the resulting distortions affect the electronic band structure, particularly the low-energy ($e_g$) subspace. To arrive at a quantitative statement, we derive effective interactions by downfolding the optimised electronic structures using the constrained random phase approximation (cRPA) \cite{aryasetiawan_frequency-dependent_2004}. 
Our results reveal a pressure-driven competition between, on the one hand, orbital localisation, reduced octahedral distortions, and overall band splitting, which enhance the interaction strengths, and, on the other hand, increased screening from the spacer La $5d_{x^{2}-y^{2}}$ bands, which dominates in the high-pressure limit. This interplay shapes the pressure dependence of the low-energy electronic correlations, which we find closely track the experimentally observed evolution of $T_\text{c}$.
To investigate the important role of the \textit{A}-site cation in this competition between interaction-increasing effects and screening, we replace La with Ac, which lowers the critical pressure but also predicts a reduced $T_\text{c}$.
Our effective parameters furthermore provide a starting point for future studies exploring the pairing mechanism in bilayer nickelates via first-principles models. Finally, this work presents, to our knowledge, the first \textit{ab initio} molecular dynamics (AIMD) simulations of La$_3$Ni$_2$O$_7$, explicitly capturing finite-temperature effects and providing new insights into the experimentally debated structural phase diagram, thereby further clarifying the origin of the right-triangular superconducting dome.

\section*{Results}
\subsection*{Pressure-Driven Structural Distortions} \label{subsec: structural optimisation}
\begin{figure}[t]
    \centering
    \includegraphics[width=0.9\linewidth]{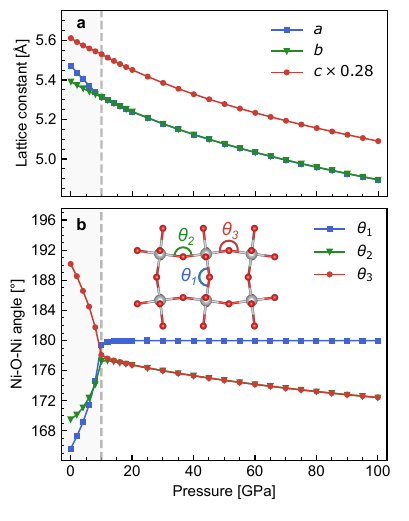}
    \caption{\textbf{Enthalpic optimisation of bilayer La$_3$Ni$_2$O$_7$.}
    \textbf{a} Lattice parameters of the orthorhombic unit cell at hydrostatic pressures between 0 and 100 GPa. 
    \textbf{b} The three inequivalent Ni-O-Ni bond angles of the $Cmcm$ phase. The two intralayer angles $\theta_2$ and $\theta_3$ in the $Cmcm$ phase (grey) become equal in the $I4/mmm$ phase (above 10 GPa), and coincidentally the NiO$_6$ octahedra align along the $c$-axis ($\theta_1=180^\circ$).} 
    \label{fig: bilayer geometries} 
\end{figure}

Fig.~\ref{fig: bilayer geometries}a shows the optimised lattice parameters of the orthorhombic unit cell of La$_3$Ni$_2$O$_7$, defined in Fig.~\ref{fig: introduction}a, at hydrostatic pressures ranging from 0 to 100 GPa. We focus on the bilayer structure, as the experimentally observed alternating monolayer-trilayer polymorph exhibits a higher enthalpy everywhere within the considered pressure range (as shown in Supplementary Fig. 1).

Two distinct regimes in both degeneracy and slopes can be observed. In the low-pressure region up to $\sim$10 GPa, the in-plane lattice constants $a$ and $b$ as depicted in Fig.~\ref{fig: introduction}c, are non-degenerate. At higher external pressures, these become equal, thereby eliminating the Jahn-Teller distortion in the $ab$-plane. This explicitly confirms the absence of a zero-temperature, high-pressure orthorhombic \textit{Fmmm} phase where $a\neq b$. Fig.~\ref{fig: bilayer geometries}b shows the three inequivalent Ni-O-Ni bond angles in the $Cmcm$ phase. The interlayer (apical) bond angle $\theta_1$ becomes equal to $180^\circ$ at the transition to the tetragonal phase around 10 GPa, indicating the alignment of the NiO$_6$ octahedra along the $c$-axis. Furthermore, the two inequivalent intralayer (basal) bond angles in the low-pressure structures, $\theta_2$ and $\theta_3$, become equal and both face inwards (\textit{i.e.} into the bilayer), which restores periodicity in the $ab$-plane, as depicted in Fig.~\ref{fig: introduction}c. The combined effect of symmetry restoration in the $ab$-plane and the alignment of the NiO$_6$ octahedra reproduces the structural transition from the low-pressure phase with space group $Cmcm$ to the high-pressure tetragonal phase with subgroup $I4/mmm$, consistent with previous work \cite{geisler_structural_2024}. The pressure dependence of this phase transition reflects the enthalpic competition between the energetically favoured, distorted $Cmcm$ cells and the $PV$ contribution, which stabilises the more compact $I4/mmm$ structure, as detailed in Supplementary Note 2. We attribute the experimental stabilisation of a zero-temperature intermediate $Fmmm$ phase to unintended anisotropic stress, as explicitly simulated by Peng \textit{et al.}~\cite{peng_importance_2024}.

In the high-pressure limit the cell contracts isotropically, as can be seen from the parallel curves corresponding to the three lattice constants in Fig.~\ref{fig: bilayer geometries}a. However, the crystal structure in the $I4/mmm$ phase compresses anisotropically as the intralayer angle $\theta_2$ in Fig.~\ref{fig: bilayer geometries}b reaches its maximum value of $\sim 177.3^\circ$ near the structural transition and then decreases monotonically with increasing pressure. This buckling effect can be explained by the La ions from the neighbouring rock-salt (spacer) layers, closing in on the bilayers along the $c$-axis as the pressure increases. In the next section, we examine the impact of the structural phase transition and this high-pressure anisotropic compression on the low-energy electronic structure.

\subsection*{Electronic Structure and Dimer Model} \label{subsec: electronic structure}
\begin{figure*}[t]
    \centering
    \includegraphics[width=\linewidth]{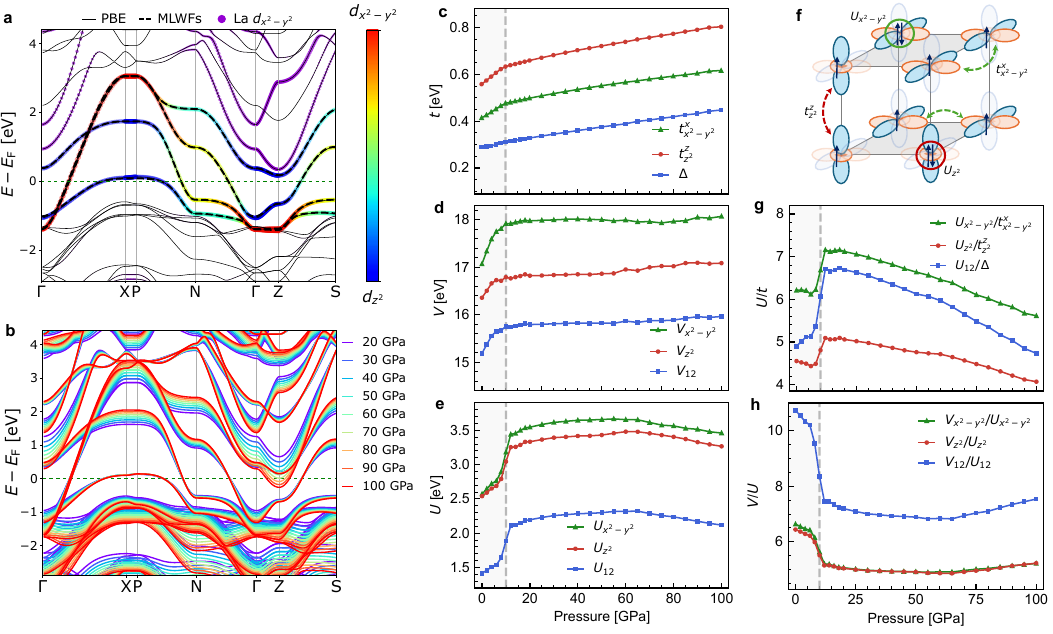}
    \caption{\new{
    \textbf{Pressure dependence of the band structure and dimer model of La$_3$Ni$_2$O$_7$.}
    \textbf{a} Band structure at 40 GPa, shown as solid lines, with the colour scale indicating the overlap with the different Ni $e_g$ orbitals. The projection onto the spacer La $d_{x^2-y^2}$ orbitals is superimposed as purple circles, with radii representing the weight. The Wannier-interpolated bands obtained from the four maximally localised Wannier functions (MLWFs) are shown as dashed lines.
    \textbf{b} Band structures at pressures between 20 and 100 GPa.
    \textbf{c} Pressure dependence of the largest hopping parameters in the dimer model: $\Delta = \epsilon (d_{x^2-y^2}) - \epsilon (d_{z^2})$ is the energy difference between MLWFs, $t^x_{x^2-y^2}$ represents nearest-neighbour intralayer hopping between $d_{x^2-y^2}$ MLWFs and $t^z_{z^2}$ represents interlayer hopping between $d_{z^2}$ MLWFs.
    \textbf{d} Bare and \textbf{e} screened on-site Coulomb interactions: $V_{x^2-y^2}$ ($U_{x^2-y^2}$) and $V_{z^2}$ ($U_{z^2}$) are the interactions in the $d_{x^2-y^2}$ and $d_{z^2}$ MLWFs as shown in \textbf{f}, while $V_{12}$ ($U_{12}$) denotes the interaction between different on-site MLWFs.
    \textbf{g} Ratios of screened interaction strengths to their associated hopping parameters and \textbf{h} bare-to-screened interaction ratios.
    }}
    \label{fig: electronic}
\end{figure*}

To understand the origin of superconductivity, it is crucial to develop an effective Hamiltonian that adequately describes the electrons near the Fermi surface. Fig.~\ref{fig: electronic}a shows the electronic band structure of La$_3$Ni$_2$O$_7$ in the $I4/mmm$ phase at 40 GPa, with energies referenced to $E_\text{F}$. The bands are projected onto the Ni $e_g$ orbitals, with the colour scale indicating the relative weight of the two components. Contributions from the spacer La $5d_{x^2 - y^2}$ orbitals are also included, highlighting the orbital character of the states near $E_\text{F}$. Additional $lm$-decompositions are provided in Supplementary Note 3. The four bands closest to $E_\text{F}$ in Fig.~\ref{fig: electronic}a exhibit clear $e_g$ character, with the individual contributions from $d_{z^2}$ and $d_{x^2-y^2}$ varying significantly across the Brillouin zone. To capture the low-energy physics of these bands, we employ the multi-scale \textit{ab initio} scheme for correlated electrons, incorporating screening from high-energy states through many-body perturbation theory \cite{aryasetiawan_frequency-dependent_2004, imada_electronic_2010}. This yields a generalised Hubbard model defined on a bilayer square lattice with two orbitals per site, commonly referred to as the dimer model. The negligible contribution from other spherical harmonics allows to disentangle the $e_g$ bands from higher-energy states using the projection method for Wannierisation of entangled bands \cite{souza_maximally_2001}. The resulting maximally localised Wannier functions (MLWFs) accurately span the low-energy subspace, as confirmed by the Wannier-interpolated bands in Fig.~\ref{fig: electronic}a.

For the high-pressure $I4/mmm$ phase, the evolution of the DFT band structure with pressure is shown in Fig.~\ref{fig: electronic}b. \new{The $e_g$ pockets, which are often examined in the context of superconductivity, show no significant pressure-induced changes in our PBE results, consistent with the findings of Jiang \textit{et al.}~\cite{jiang_theory_2025}.} We see that most (high-energy) bands tend to move away from $E_\text{F}$ under increasing pressure, while the dimer $e_g$ and spacer La 5$d_{x^2-y^2}$ bands become more dispersive as reported in earlier work \cite{jiang_theory_2025}. Remarkably, near the $\Gamma$ and Z points, we observe that above 45 GPa, the lowest spacer La 5$d_{x^2 - y^2}$ conduction band moves below the $e_g$ bands. Above 75 GPa, it even drops below $E_\text{F}$, thereby forming an additional electron pocket. As we will show next, this lowering of the spacer La bands has a significant effect on the effective low-energy physics.  \new{Notably, $GW$ calculations indicate that self-energy corrections, in addition to the pressure-driven lowering, further lower the spacer La $5d_{x^2-y^2}$ bands relative to PBE, pushing them below $E_\text{F}$ at pressures above $\sim$30 GPa \cite{you_unlikelihood_2025, vaulx_pressure_2025}.} For completeness, the pressure-dependence of the DFT band structure in the $Cmcm$ phase is shown in Supplementary Fig. \new{3}.

Localised model parameters are now obtained by downfolding the electronic structure onto the corresponding MLWFs. By doing this for every optimised crystal structure, Fig.~\ref{fig: electronic}\new{c,d,e} shows the pressure dependence of the most significant parameters in the dimer model (Fig.~\ref{fig: electronic}f). Specifically, we plot the three largest screened on-site interactions, $U_{x^2 - y^2}$, $U_{z^2}$, and $U_{12}$, along with their bare Coulomb counterparts $V_{x^2 - y^2}$, $V_{z^2}$, and $V_{12}$. Here, $U_{x^2 - y^2}$ and $U_{z^2}$ are the effective interactions between two electrons in MLWFs with $d_{x^2 - y^2}$ and $d_{z^2}$ character, respectively, while $U_{12}$ denotes the effective interaction between both orbitals on the same site. In addition, we show the three largest one-body terms: the orbital energy difference $\Delta = \epsilon(d_{x^2 - y^2}) - \epsilon(d_{z^2})$, the nearest-neighbour intralayer hopping $t^x_{x^2 - y^2}$, and the interlayer hopping $t^z_{z^2}$. \new{To reduce the impact of potential absolute errors in the cRPA-derived interaction parameters, we focus on relative, pressure-driven trends.}

We see that the on-site repulsion energies, $U_{x^2-y^2}, U_{z^2}$ and $U_{12}$ in Fig.~\ref{fig: electronic}e, are significantly enhanced (up to $\sim 30\%$) with the structural phase transition near 10 GPa, continue to increase and reach a maximum around 55--65 GPa, and gradually decrease with cell compression at higher pressures. In contrast, the kinetic terms, $\Delta$, $t^x_{x^2-y^2}$ and $t^z_{z^2}$ increase monotonically, yet at a slower rate than in the $Cmcm$ phase. As a measure of correlation strength in the dimer model, various $U_i/t_i$ ratios are given in Fig.~\ref{fig: electronic}g. These interaction-to-hopping ratios are abruptly enhanced near the structural phase transition, peaking around 18 GPa before gradually decreasing. Interestingly, the parameters near the structural transition at 10 GPa are nearly equivalent (up to a global scale) to those around 45-55 GPa. From the perspective of this effective dimer model, these similar electronic conditions can be qualitatively associated with the critical pressures of 14 and 51 GPa, between which the strongest signs of zero resistance have been observed \cite{li_identification_2025}. While the DFT approximation and the simplified nature of the dimer model may shift the zero temperature critical pressures, the overall trend is robust: the smaller $U_i/t_i$ ratios in both the low-pressure $Cmcm$ phase and at extreme pressures clearly signal a weakening of electronic correlations. \new{Furthermore, within this critical pressure window, the steep enhancement of the $U_i/t_i$ ratios at the structural phase transition, followed by a gradual decrease at higher pressures, closely tracks the experimentally observed phase boundary of the right-triangular superconducting region in the PT phase diagram reported by Li \textit{et al.}~\cite{li_identification_2025}. Notably, the maximum of $U_i/t_i$ in our simulations occurs at 18 GPa, coinciding with the experimentally observed maximum $T_\mathrm{c}$.} 

We can understand the pressure-dependence of the dimer model parameters as follows. In the low-pressure limit, the $Cmcm$ phase hosts a checkerboard pattern with two distinct tilted NiO$_6$ octahedra (see Fig.~\ref{fig: introduction}c). Consequently, the nearest-neighbour hopping integrals are weakened due to the tilted bond angles which reduce orbital overlap. The reduced symmetry also separates the intralayer hopping into two components: one corresponding to the inward-facing and one for the outward-facing angles. However, the difference between these parameters is minimal, remaining below $2\%$.

At low pressures, the pressure-driven enhancement of $U_i$ arises from three effects. First, the reduced overlap between low- and high-energy orbitals, resulting from the suppression of octahedral distortions, weakens screening. Second, the high-energy bands exhibit an overall tendency to move away from the Fermi level, as shown in Fig.~\ref{fig: electronic}b, which typically diminishes their ability to screen the Coulomb interaction \cite{panda_pressure_2017, tomczak_realistic_2010, kim_strain-induced_2018, ivashko_strain-engineering_2019}. Together, these effects reduce the screening strength with increasing pressure, as reflected by the strong decrease in $V_i/U_i$ in Fig.~\ref{fig: electronic}h. However, the increase in $U_i$ cannot be entirely attributed to screening effects, since the bare interactions $V_i$ also increase with pressure in the $Cmcm$ phase (see Fig.~\ref{fig: electronic}d). This third effect can be explained by the crystal compression, which leads to more localised MLWFs and consequently enhances the bare on-site interactions with pressure.

In the high-pressure limit, the compression of the crystal structure in the $I4/mmm$ phase (Fig.~\ref{fig: bilayer geometries}) enhances the overlap between nearby MLWFs, which monotonically increases the hopping integrals in Fig.~\ref{fig: electronic}\new{c}. The compression also decreases the MLWF spreads, which leads to increasing values of $\Delta$. Above 60 GPa, the screening in Fig.~\ref{fig: electronic}\new{h} increases with pressure, and the effective interaction parameters in Fig.~\ref{fig: electronic}e start to decrease. We attribute this enhanced screening to the spacer La cations, whose 5$d_{x^2-y^2}$ conduction bands energetically enclose the low-energy bands near the $\Gamma$ and Z points (Fig.~\ref{fig: electronic}b), and which spatially enclose the $e_g$ MLWFs, as indicated by the decreasing intralayer angle $\theta_2$ in Fig.~\ref{fig: bilayer geometries}b. Furthermore, the additional lowering of the spacer La $5d_{x^2-y^2}$ bands due to self-energy corrections suggests that these bands exert an even stronger screening effect on the dimer subspace \cite{you_unlikelihood_2025, vaulx_pressure_2025}. The pressure-driven competition between, on the one hand, MLWF compression, reduced octahedral distortions and band splitting, which enhance the dimer-model hopping integrals and interaction strengths, and, on the other hand, the increased screening arising from the spacer La $5d_{x^{2}-y^{2}}$ bands, which reduces the effective interactions and dominates in the high-pressure limit, shapes the $U/t$ trends discussed above. The separate trends are further substantiated by decomposing the overall screening into channels as detailed in Supplementary Note 4.

\subsection*{Finite-Temperature Effects} \label{subsec: aimd}

\begin{figure*}[t]
    \centering
    \includegraphics[width=\linewidth]{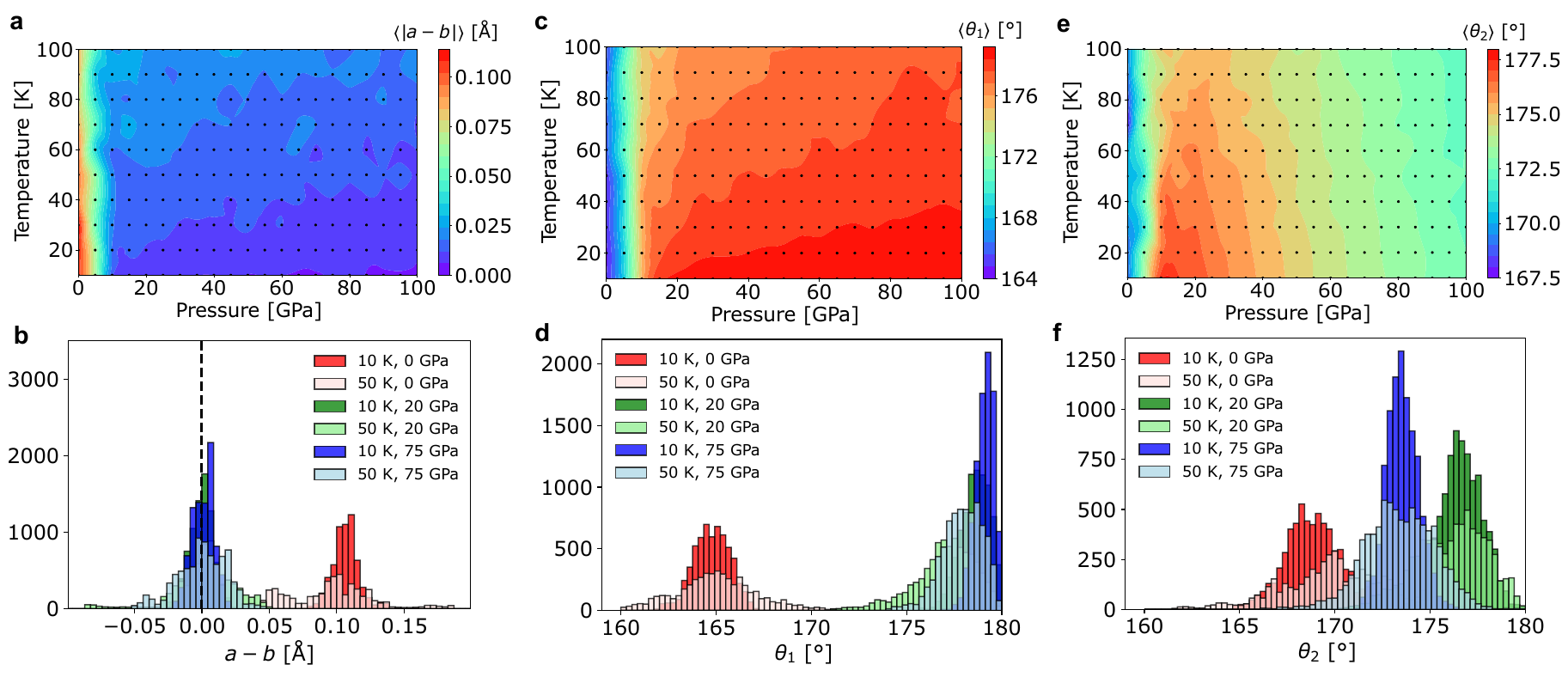}
    \caption{\textbf{AIMD simulations of La$_3$Ni$_2$O$_7$ in the range of 10–100 K and 0–100 GPa.} 
    The upper row shows the time-averages of the full diagram, \new{with black dots marking the simulated sampling points}. The lower row displays frequency histograms at six pressure-temperature points.  
    Panels \textbf{(a,b)} show the difference between in-plane lattice constants $a$ and $b$, \textbf{(c,d)} the interlayer angle $\theta_1$, and \textbf{(e,f)} the inwards intralayer angle $\theta_2$.
    }
    \label{fig: AIMD}
\end{figure*}

Next we include finite temperature (10–100 K) effects through AIMD simulations, extending pressure-driven structural and electronic trends along the temperature axis. Fig.~\ref{fig: AIMD}a shows the time-averaged absolute difference between the in-plane lattice constants $a$ and $b$. This is a measure for the finite-temperature fluctuations in $a-b$ away from zero. We see that $\langle |a - b| \rangle$ is on the order of $\sim$0.1 Å at low temperatures and low pressures, corresponding to the orthorhombic $Cmcm$ phase. Upon increasing pressure $\langle |a-b|\rangle$ quickly drops to $\sim$ 0.01 Å in the tetragonal $I4/mmm$ phase. The region of rapid decrease in $\langle |a-b|\rangle$ extends almost vertically in the PT-phase diagram, with a slope of approximately -75 K/GPa, up to the maximum temperature of $100$ K used in this work. \new{This observation aligns with the negative slope initially reported experimentally by Wang \textit{et al.}~\cite{wang_temperature-dependent_2025} and is close to the slope of -60 K/GPa reported by Asai \textit{et al.}~\cite{asai_finite-temperature_2025} based on finite-temperature structural optimisations that include anharmonic vibrational contributions to the free energy.} Fig.~\ref{fig: AIMD}a furthermore implies that\new{, if present,} the temperature-driven transition from the high-pressure $I4/mmm$ tetragonal phase found in DFT to the orthorhombic $Fmmm$ phase observed in XRD at room temperature must occur at temperatures above $100$ K, placing this transition at a comfortable distance from the superconducting region. This can be further confirmed by looking at $a-b$ histograms, as shown for six PT-points in Fig.~\ref{fig: AIMD}b. In the region of large $\langle |a-b|\rangle$ at low pressure (exemplified at 0 GPa), we see that the $a-b$ histograms are centred around $a-b \approx$ 0.1 Å, both at $T=10$ K and $T=50$ K. For higher pressures, on the other hand, the $a-b$ histograms are clearly centred around $0$ at both $T=10$ K and $T=50$ K. \new{To probe the possible $Fmmm$ phase at higher temperatures, we performed additional AIMD simulations at 300 K (see Supplementary Note 5). No evidence for this phase was observed, and we again conclude that its experimental stabilisation may arise from anisotropic pressure.}

The time-averaged interlayer angle $\theta_1$ is shown in Fig.~\ref{fig: AIMD}c. Note that $\theta_1$ is defined as the internal bond angle, with $\theta_1  = 180^\circ$ representing a fully straightened bond and $\langle \theta_1\rangle$ a measure for the fluctuations away from $\theta_1=180^\circ$. We see from Fig.~\ref{fig: AIMD}c that $\langle \theta_1\rangle$ exhibits two distinct trends. First, approximately independent of temperature, $\langle\theta_1\rangle$ quickly increases from $\sim 165^\circ$ to $\sim 180^\circ$ around a pressure of $10$ GPa. This we attribute to the octahedral tilt which disappears upon going from $Cmcm$ to $I4/mmm$, which can be verified explicitly from the histograms in Fig.~\ref{fig: AIMD}d: the low-pressure histograms are centred around $\sim 165^\circ$ while the high-pressure histograms are centred around $\sim 180^\circ$. The reduced bin heights near $\theta_1=180^\circ$ reflects the unscaled space angle, resulting from azimuthal freedom. The verticality of the region where $\langle \theta_1\rangle$ quickly rises agrees with the structural transition discussed above. This result indicates that the experimentally observed superconducting region is contained entirely within the $I4/mmm$ phase. Secondly, inside the $I4/mmm$ phase, increasing the temperature reduces $\langle \theta_1\rangle$ due to enhanced thermal fluctuations. At pressures right after the structural transition ($14-20$ GPa), we see that increasing the temperature significantly affects the interlayer angle. Higher hydrostatic pressure, in turn, enhances the robustness of octahedral alignment against thermal fluctuations.

Also $\langle \theta_2\rangle$, the average of the intralayer Ni-O-Ni bond angle, exhibits a narrow region of rapid increase at the $Cmcm$ -- $I4/mmm$ phase boundary, as shown in Fig.~\ref{fig: AIMD}e. At higher pressures, we see that crystal compression leads to a decreasing value of $\langle \theta_2\rangle$, due to the La ions approaching the NiO layers along the $c$-axis as observed at 0 K in Fig.~\ref{fig: bilayer geometries}b. Increasing temperature also reduces $\langle \theta_2\rangle$. From the histograms in Fig.~\ref{fig: AIMD}f we see that $\theta_2$ is sufficiently far away from $180^\circ$ to be able to conclude that the decrease in $\langle \theta_2\rangle$ is not a fluctuation effect, but indicates a true shift in average value of the intralayer bond angle.


\new{These finite-temperature structural variations can now be leveraged to further strengthen the link between electronic correlations and the experimentally observed right-triangular superconducting region in the PT-phase diagram \cite{li_identification_2025}. At the lower critical pressure, the rapid increase of correlations in Fig.~\ref{fig: electronic}\new{g}, and consequently of $T_\text{c}$, remains unaffected by the nearly vertical $Cmcm$ - $I4/mmm$ phase boundary, where the octahedral tilt vanishes. At higher pressures, the pressure-driven robustness of $\theta_1$ and $|a-b|$ to thermal fluctuations helps maintain these structural conditions favourable for superconductivity. In contrast, the buckling of the intralayer angle $\theta_2$ is further enhanced with both pressure and temperature, leading to stronger screening by the spacer La cations, thereby weakening correlations. This finite-temperature effect may contribute to the observed suppression of $T_\mathrm{c}$ under increased pressure.}

\subsection*{\new{\textit{A}-site Cation Substitution}} \label{subsec: chemical substitution}

\begin{figure*}[t] 
    \centering
    \includegraphics[width=\linewidth]{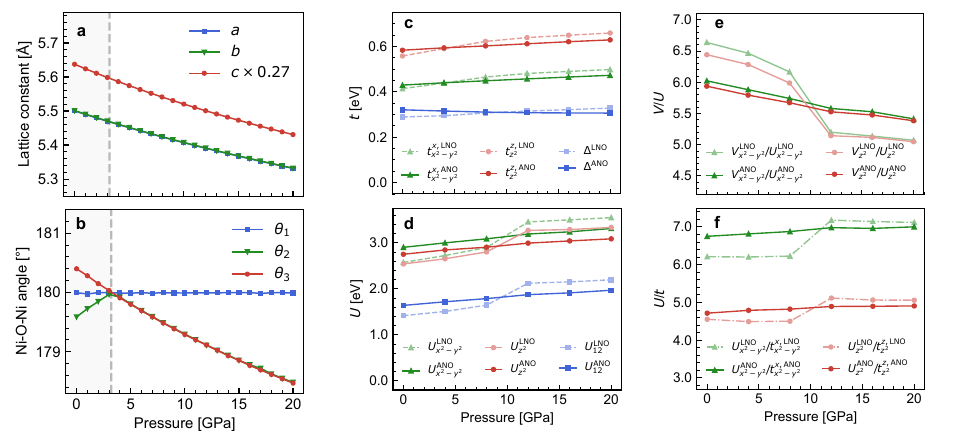}
    \caption{\new{\textbf{Pressure-dependent structural and dimer model properties of Ac$_3$Ni$_2$O$_7$ between 0 and 20 GPa.}
    \textbf{a} Lattice constants of the orthorhombic unit cell (as shown for La$_3$Ni$_2$O$_7$ in Fig.~\ref{fig: introduction}a).
    \textbf{b} The three unique Ni-O-Ni bond angles of the $Cmcm$ phase.
    \textbf{c} Hopping parameters and \textbf{d} screened interaction strengths in the dimer model as defined for La$_3$Ni$_2$O$_7$ in Fig.~\ref{fig: electronic}. Superscripts denote the material: ANO for Ac$_3$Ni$_2$O$_7$ and LNO for La$_3$Ni$_2$O$_7$. \textbf{e} Bare-to-screened interaction ratios and \textbf{f} ratios of screened interaction strengths to their associated hopping parameters.}
    } 
    \label{fig: ANO}
\end{figure*}

Our analysis of bilayer La$_3$Ni$_2$O$_7$ can now be used as a guiding principle for the design of bulk nickelates superconducting near ambient pressure. The most straightforward adaptation is an isoelectronic substitution of the rare-earth $A$-site cation La$^{3+}$, which plays a pivotal role in the structural phase transition and corresponding critical pressure. Doping with smaller atoms was initially expected to stabilise the tetragonal phase at lower pressures via chemical pre-compression. However, DFT simulations on $A_3$Ni$_2$O$_7$ showed the opposite: smaller atoms increased the transition pressure from the orthorhombic to the tetragonal structure \cite{geisler_structural_2024}. Consequently, larger substituents were proposed to stabilise the $I4/mmm$ structure by the inherent chemical pressure, replacing external hydrostatic pressure. Among (valence) isoelectronic elements, only Ac$^{3+}$ possesses a larger ionic radius, for which Ac$_3$Ni$_2$O$_7$ was indeed found to be structurally stable in the $I4/mmm$ phase at ambient pressure, as confirmed by DFT+$U$ calculations \cite{rhodes_structural_2024}.

Enthalpic optimisation of bilayer Ac$_3$Ni$_2$O$_7$ at external pressures between 0 and 20 GPa, as shown in Fig.~\ref{fig: ANO}\new{a,b}, confirms the absence of Jahn-Teller distortion in the $ab$-plane, which is likely due to the increased chemical pressure of the larger Ac$^{3+}$ cations. However, we find that the intralayer spatial symmetry is still (slightly) broken at pressures up to 3.2 GPa. The two inequivalent intralayer angles $\theta_2$ and $\theta_3$ in Fig.~\ref{fig: ANO}b become equal with NiO$_6$ octahedral alignment in the $I4/mmm$ phase above 3.2 GPa. The presence of a low-pressure $Cmcm$ phase could arise from limitations in the PBE-D3 level, as opposed to the DFT+$U$ methodology used in Ref.~\cite{rhodes_structural_2024}, which takes more electronic correlations into account and found an $I4/mmm$ phase already at $0$ GPa. Nonetheless, the significantly lower critical pressure found in this work, resulting from the enlarged ionic radius, confirms the important role of chemical pressure exerted by the $A$-site cations.

\new{Beyond structural changes, the effect of this chemical substitution on the low-energy correlations must be examined to understand its impact on superconductivity. Since the band structures of La$_3$Ni$_2$O$_7$ and Ac$_3$Ni$_2$O$_7$ are nearly identical \cite{rhodes_structural_2024}, their effective dimer models can be directly compared. Wu \textit{et al.}~\cite{wu_ac_3ni_2o_7_2024} first noted that La$_3$Ni$_2$O$_7$ at 30 GPa and Ac$_3$Ni$_2$O$_7$ at ambient pressure exhibit nearly identical hopping integrals and screened interactions, supporting the possibility of ambient-pressure superconductivity in Ac$_3$Ni$_2$O$_7$. We extend their analysis by calculating the dimer model parameters from 0 to 20 GPa, revealing the pressure-dependent effects of this substitution (Fig.~\ref{fig: ANO}c,d). While Ac$_3$Ni$_2$O$_7$ exhibits larger values of both $U$ and $t$ at ambient pressure, the structural phase transition in La$_3$Ni$_2$O$_7$ leads to a reversal of this trend. The lower effective interactions at high pressures for Ac$_3$Ni$_2$O$_7$ are a result of the enhanced screening by Ac as seen from the bare-to-screened interaction ratios, $V_i/U_i$, in Fig.~\ref{fig: ANO}e. A more direct comparison of the low-energy physics is obtained by examining the ratios $U_i/t_i$, as shown in Fig.~\ref{fig: ANO}f. For pressures below 10 GPa, Ac$_3$Ni$_2$O$_7$ exhibits notably stronger correlations than La$_3$Ni$_2$O$_7$. At ambient pressure, the $U/t$ values of Ac$_3$Ni$_2$O$_7$ are comparable to those of La$_3$Ni$_2$O$_7$ at 10 and 45–55 GPa, which we associate in Fig.~\ref{fig: electronic}\new{g} with the experimental critical pressures \cite{li_identification_2025}. This similarity suggests that Ac$_3$Ni$_2$O$_7$ may exhibit superconductivity near ambient pressure, although it more strongly indicates that the onset could occur slightly above ambient pressure, yet still below the critical pressure of the La variant. However, the overall lower $U_i/t_i$ values of Ac$_3$Ni$_2$O$_7$ compared with the maximal values in La$_3$Ni$_2$O$_7$ indicate a reduced maximal $T_\mathrm{c}$. This is consistent with a recent machine learning study by Li \textit{et al.}~\cite{li_machine_2025}, which predicts a $T_\mathrm{c}$ of 70.3 K for Ac$_3$Ni$_2$O$_7$ versus 80 K for La$_3$Ni$_2$O$_7$.}

\section*{Methods} \label{sec: Methodology}
\paragraph*{\textbf{Electronic Structure Calculations}}
The first-principles DFT calculations are performed using the Vienna Ab initio Simulation Package (VASP) employing the Perdew-Burke-Ernzerhof (PBE) exchange-correlation functional and the projector-augmented-wave \new{(PAW)} formalism with a plane-wave energy cut-off of 600 eV  \cite{kresse_ab_1993, kresse_efficient_1996, perdew_generalized_1996, blochl_projector_1994}. \new{The valence electron configurations of the PAW potentials are La($4f^{0}$$5s^2$$5p^6$$5d^{1}$$6s^2$), Ac($6s^{2}$$6p^6$$6d^1$$7s^2$), Ni($3d^9$$4s^1$) and O($2s^2$$2p^4$). The La PAW potential explicitly includes the unoccupied $4f$ states, which at the PBE level form a set of narrow conduction bands and may spuriously hybridise with low-energy states. As discussed in detail in Supplementary Note 6, a Hubbard correction of $U = 18$ eV was therefore applied to the La $4f$ orbitals in order to shift these states well above the La $5d$ manifold, fully disentangle them from the low-energy subspace, and eliminate non-physical hybridisation.} \new{Hubbard $U$ corrections are, however, not applied for structural optimisations and AIMD simulations because the atomic-orbital projection scheme (as used in VASP) can introduce nonphysical fractional $d$- and $f$-orbital occupations, leading to spurious forces \cite{warda_getting_2026}.} All calculations are performed in the non-magnetic (spin-unpolarised) case. The convergence criterion for the electronic self-consistent loops is set to 10$^{-8}$ eV. Brillouin-zone sampling is $\Gamma$-centred, where convergence is systematically validated. \\

\paragraph*{\textbf{Structural Optimisation}}
Structural optimisations are performed at fixed hydrostatic pressures by minimising the enthalpy using the conjugate gradient algorithm with respect to all ionic and cell degrees of freedom until the change in total energy subceeds 10$^{-7}$ eV and all Hellmann-Feynman forces subceed 10 meV/\AA. To avoid finite size effects, charge densities are computed on a 9$\times$9$\times$3 $k$-point grid for the largest unit cell among all experimentally known phases, with convergence validated In Supplementary Note 7. Initial cells contain distorted Wyckoff positions from experimental data \cite{sun_signatures_2023}. Pairwise van der Waals dispersion corrections are included via the DFT-D3 method with Becke-Johnson damping as justified in Supplementary Note 8 \cite{grimme_consistent_2010, grimme_effect_2011} \\

\paragraph*{\textbf{Downfolding}}
The single-point electronic structures are recalculated on a $\Gamma$-centred 5$\times$5$\times$9 $k$-point grid for the primitive unit cells in the $Cmcm$ phase. The resulting 336 lowest bands, respectively, are then downfolded into the $e_g$ dimer configuration using the multi-scale \textit{ab initio} scheme for correlated electrons developed by Imada \textit{et al.}~\cite{imada_electronic_2010}. Within this scheme, we employ the projector cRPA disentanglement method and tight-binding methodology to obtain the screened Coulomb interactions and hopping integrals, respectively \cite{aryasetiawan_frequency-dependent_2004,  projector_Hanke_Sham, kaltak_projector}. Comparisons to other disentanglement schemes are discussed in Supplementary Note 3 to avoid artifacts associated with this particular choice. Static interactions were assumed, as frequency dependence from retarded screening was found to be negligible (see Supplementary Note 3). To construct a sufficiently localised basis spanning this low-energy subspace, MLWFs are constructed from projecting the Kohn-Sham orbitals on the $e_g$ orbitals \cite{marzari_maximally_localized_1997, pizzi_wannier90_2020, souza_maximally_2001}. The dispersion of the Wannier-interpolated bands is systematically compared to the DFT bands, confirming a faithful disentanglement. \\

\paragraph*{\textbf{Molecular Dynamics}}
The AIMD simulations are conducted on a 3$\times$3$\times$1 $k$-point grid for the orthorhombic unit cell (Fig.~\ref{fig: introduction}a) in the isothermal-isobaric (NPT) ensemble, using the Langevin thermostat and the Parrinello-Rahman barostat as implemented in VASP \cite{allen2017computer, parrinello_crystal_1980, parrinello_polymorphic_1981}. A barostat mass of 2000 amu was employed to maintain stable pressure control. Langevin damping coefficients of 10 ps$^{-1}$, 5 ps$^{-1}$, and 2 ps$^{-1}$ were applied to the La, Ni, and O atoms, respectively, while a lattice damping coefficient of 10 ps$^{-1}$ was used. A time step of 1 fs was adopted, with initial configurations equilibrated for 2 ps prior to data collection. The phase diagram was interpolated from a square grid comprising 21 pressure values (0–100 GPa in steps of 5 GPa) and 10 temperature values (10–100 K in steps of 10 K).

\vspace{2em}

\new{
\section*{Data availability}
Source Data for Figs.~\ref{fig: bilayer geometries}, \ref{fig: electronic}c,d,e, \ref{fig: AIMD} and \ref{fig: ANO} are provided with this paper. Any additional data that support the findings of this study are available from the corresponding author upon request.}

\new{
\section*{Code availability}
All VASP input data and post-processing code supporting the results of this work are available from the online GitHub repository at https://github.com/daanverraes/Supporting-Information or upon request from the authors. Pseudocode outlining the custom post-processing scripts for both the Wannier-interpolated band structures and the calculation of tight-binding hopping parameters is provided in Supplementary Note 3.}

\section*{Acknowledgments}
This work is supported by the Research Board of Ghent University (BOF) through a Concerted Research Action (BOF23/GOA/021). V. V. S. acknowledges the Research Board of Ghent University (BOF). The computational resources (Stevin Supercomputer Infrastructure) and services used in this work were provided by VSC (Flemish Supercomputer Center), funded by Ghent University, FWO, and the Flemish Government—department EWI. N. B. was supported by the European Research Council under the European Union Horizon 2020 Research and Innovation Programme via Grant Agreement No. 101076597-SIESS, and by a grant from the Simons Foundation (SFI-MPS-NFS-00006741-04).

\section*{Author contributions}
D. V., T. B., N. B. and V. V. S. initiated the discussion, designed the paper, and were involved in the discussion of the results. D. V. and N. B. wrote the manuscript with contributions of all authors. D. V. performed all simulations.

\section*{Competing Interests}
The authors declare no competing interests.

\end{document}